\documentclass[conference]{IEEEtran}

\ifCLASSINFOpdf

\else

\fi

\usepackage{amsmath}

\usepackage{tikz}
\usepackage{pgfplots}
\usepackage{pgf}
\usepackage{amssymb}
\usepackage{amsthm}
\newtheorem{theorem}{Theorem}

\usepackage[ruled,vlined]{algorithm2e}

\definecolor{myblue}{rgb}{0.00000,0.44700,0.74100}
\definecolor{myred}{rgb}{0.85000,0.32500,0.09800}
\definecolor{mycolor3}{rgb}{0.92900,0.69400,0.12500}
\definecolor{mypurple}{rgb}{0.49400,0.18400,0.55600}
\definecolor{mygreen}{rgb}{0.46600,0.67400,0.18800}

\begin{document}

\title{On Time-Bandwidth Product of Multi-Soliton Pulses\hspace*{-1pt}\vspace*{-0.3ex}}

\author{\IEEEauthorblockN{Alexander Span$^*$, Vahid Aref$^\dag$, Henning B\"ulow$^\dag$, and Stephan ten Brink$^*$}
\IEEEauthorblockA{$^*$Institute of Telecommunications, University of Stuttgart, Stuttgart, Germany\\
		$^\dag$Nokia Bell Labs, Stuttgart, Germany\vspace*{-0.1ex}
}
}

\maketitle

\begin{abstract}
Multi-soliton pulses are potential candidates for fiber optical transmission
where the information is modulated and recovered in the so-called nonlinear Fourier 
domain.
While this is an elegant technique to account for the channel nonlinearity,
the obtained spectral efficiency, so far, is not competitive with the classic Nyquist-based schemes. In this paper, we study the evolution of the time-bandwidth product of multi-solitons as they propagate along the optical fiber. 
For second and third order soliton pulses, we numerically optimize the pulse shapes to 
achieve the smallest time-bandwidth product when the phase of the spectral amplitudes
is used for modulation.
Moreover, we analytically estimate the pulse-duration and bandwidth of multi-solitons in some practically important cases. Those estimations enable us to approximate the time-bandwidth product for higher order solitons.
\end{abstract}

\IEEEpeerreviewmaketitle

\section{Introduction}

Advances made over the past decade in coherent optical technology have significantly improved transmission capacities to a point where Kerr nonlinearity once again becomes the limiting factor. 
The equalization of nonlinear effects is usually very complex and has a limited gain due to the mixing of signal and noise on the channel.
The optical channel is usually modeled by the Nonlinear Schr{\"o}dinger Equation (NLSE) 
which describes the interplay between Kerr nonlinearity and chromatic dispersion along the fiber.

The Nonlinear Fourier Transform (NFT) is a potential way of generating pulses 
matched to a channel governed by the NLSE. It maps a pulse to the nonlinear Fourier spectrum
with some beneficial properties.
This elegant technique, known also as inverse scattering method~\cite{shabat1972exact}, 
has found applications in fiber optics when the on-off keying of first order solitons 
was developed in the 1970s~\cite{mollenauer2006solitons}. Following~\cite{Yousefi2014nft,prilepsky2014nonlinear}, it has regained attention as coherent technology allows to exploit all degrees of freedom offered by the nonlinear spectrum.
 
Multi-soliton pulses are specific solutions of the NLSE. Using the NFT, an $N-$th order soliton, denoted here by $N-$soliton, is 
mapped to a set of $N$ distinct nonlinear frequencies, called eigenvalues, and the corresponding spectral amplitudes.
The key advantage of this representation is that
the complex pulse evolution along the fiber
can be expressed in terms of spectral amplitudes which evolve linearly in the nonlinear spectrum. Moreover, the transformation is \emph{independent} of the other spectral amplitudes and eigenvalues.
These properties motivate to modulate data using spectral amplitudes.
 
On-off keying of 1-soliton pulses, also called fundamental solitons, has been intensively 
studied two decades ago for different optical applications (see \cite{mollenauer2006solitons} and reference therein). To increase spectral efficiency, it has been proposed to modulate multi-solitons~\cite{Yousefi2014nft}. One possibility is the independent on-off keying of 
$N$ predefined eigenvalues. The concept has been experimentally shown up to using 10 eigenvalues
in \cite{dong2015nonlinear}, \cite{aref2016onoff}. The other possibility is to modulate the 
spectral amplitudes of $N$ eigenvalues. 
The QPSK modulation of spectral amplitudes has been verified experimentally
up to 7 eigenvalues in~\cite{aref2015experimental,Buelow20167eigenvalues,geisler2016nfdm}. All of these works have a small spectral efficiency.

Characterizing the spectral efficiency of multi-soliton pulses is still an open problem. 
First, the statistics of noisy received pulses in the nonlinear spectrum have not yet been fully understood, even though
there are insightful studies for some special cases and under some assumptions~\cite{derevyanko2005fokker,zhang2015spectral,Wahls2017noise}. Second, 
the bandwidth and the pulse-duration change as a multi-soliton propagates along a fiber or as spectral amplitudes are modulated.
The nonlinear evolution makes it hard to estimate the time-bandwidth product of a multi-soliton.

In this paper, we study the evolution of pulse-duration and bandwidth of multi-soliton pulses along an optical fiber link. We numerically optimize the time-bandwidth product of $N-$soliton pulses for $N=2$ and $3$. The results provide some guidelines for $N>3$.
We focus on scenarios where the phases of $N$ spectral amplitudes are modulated independently. However, our results can also be applied to 
on-off keying modulation schemes. 
We assume that the link is long enough 
so that the pulse-duration and bandwidth can reach their respective maximum. We also neglect inter-symbol interference.
Our results show that the optimization of \cite{hari2016multieigenvalue} 
is suboptimal when the evolution along the fiber is taken into account.

We further introduce a class of $N-$solitons which are provably symmetric. 
A subset of these pulses are already used in \cite{aref2015experimental,Buelow20167eigenvalues,hari2016multieigenvalue}.
We derive
an analytic approximation of their pulse-duration. Numerical observations exhibit that
the approximation is tight and can serve as a lower-bound 
for other $N-$solitons. 
To the best of our knowledge, this is the first result on the pulse-duration of 
multi-solitons. We also approximate the time-bandwidth product by lower-bounding the maximal bandwidth.

\section{Preliminaries on Multi-Soliton Pulses}\label{sec:pre}
In this section, we briefly explain the nonlinear Fourier transform (NFT), the characterization of multi-soliton pulses in the corresponding nonlinear spectrum and how they can be generated via the inverse NFT. 
\subsection{Nonlinear Fourier Transform}
The pulse propagation along an ideally lossless and noiseless fiber is characterized using the standard Nonlinear Schr{\"o}dinger Equation (NLSE)
\begin{equation}
\frac{\partial}{\partial z}q(t,z)+j\frac{\partial^2}{\partial t^2}q(t,z)+2j|q(t,z)|^2q(t,z)=0.
\label{NLSE}
\end{equation}
The physical pulse $Q(\tau,\ell)$ at location $\ell$ along the fiber is then described by
\begin{equation*}
Q\left(\tau,\ell\right)=\sqrt{P_0}\,\,\, q\left(\frac{\tau}{T_0},\ell \frac{\left|\beta_2\right|}{2T_0^2}\right) \text{ with } P_0\cdot T_0^2=\frac{\left|\beta_2\right|}{\gamma}, 
\end{equation*}
where $\beta_2<0$ is the chromatic dispersion and $\gamma$ is the Kerr nonlinearity of the fiber, and $T_0$ determines the symbol rate. The closed-form solutions of the NLSE \eqref{NLSE} can be described in a nonlinear spectrum defined by the following so-called Zakharov-Shabat system~\cite{shabat1972exact}
\begin{equation}\label{eq:ZS}
\frac{\partial }{\partial t}\left(\begin{matrix}\vartheta_1(t;z)\\\vartheta_2(t;z)\end{matrix}\right)=
	\left(\begin{matrix}
	-j\lambda & q\left(t,z\right) \\-q^*\left(t,z\right) & j\lambda
	\end{matrix}\right)
	\left(\begin{matrix}\vartheta_1(t;z)\\\vartheta_2(t;z)\end{matrix}\right),
\end{equation}
with the boundary condition
\[
\left(\begin{matrix}\vartheta_1(t;z)\\\vartheta_2(t;z)\end{matrix}\right)\to\left(\begin{matrix}1\\0\end{matrix}\right)\exp\left(-j\lambda t\right) 
\text{ for }  
t\to -\infty
\]
under the assumption that $q(t;z)\to 0$ decays sufficiently fast as $|t|\to \infty$ (faster than any polynomial).
The nonlinear Fourier coefficients (Jost coefficients) are defined as
\begin{align*}
a\left(\lambda;z\right) & =\lim_{t\to \infty}\vartheta_1(t;z)\exp\left(j\lambda t\right)
\\ b\left(\lambda;z\right) & =\lim_{t\to \infty}\vartheta_2(t;z)\exp\left(-j\lambda t\right).
\end{align*}
The set $\Omega$ denotes the set of simple roots of $a(\lambda;z)$ with positive 
imaginary part, which are called \textit{eigenvalues} as they do not change in terms of $z$, i.e. $\lambda_k(z)=\lambda_k$. The nonlinear spectrum is usually described by the following two parts:

\begin{itemize}
\item[(i)] Continuous Part: the spectral amplitude $Q_c(\lambda;z)=b(\lambda;z)/a(\lambda;z)$ for real frequencies $\lambda\in\mathbb{R}$.
\item[(ii)] Discrete Part: $\{\lambda_k,Q_d(\lambda_k;z)\}$ where 
$\lambda_k\in\Omega$, i.e. $a(\lambda_k;z)=0$, and
$Q_d(\lambda_k;z)=b(\lambda_k;z)/\frac{\partial a(\lambda;z)}{\partial \lambda}|_{\lambda=\lambda_k}$.
\end{itemize}
An $N-$soliton pulse is described by the discrete part only and the continuous part is equal to zero (for any $z$). 
The discrete part contains $N$ pairs of eigenvalue and the corresponding spectral amplitude, i.e. $\{\lambda_k,Q_d(\lambda_k;z)\},1\leq k\leq N $. 

An important property of the nonlinear spectrum is its simple linear evolution given by~\cite{Yousefi2014nft}
\begin{equation}\label{eq:qd_evol}
Q_d(\lambda_k;z)=Q_d(\lambda_k)\exp(-4j\lambda_k^2z),
\end{equation} 
where we define $Q_d(\lambda_k)=Q_d(\lambda_k;z=0)$. The transformation is linear and 
depends only on its own eigenvalue $\lambda_k$. This property motivates for modulation of data
over independently evolving spectral amplitudes.

Note that there are several methods to compute the nonlinear spectrum by numerically solving  
the Zakharov-Shabat system. Some of these methods are summarized in \cite{Yousefi2014nft},\cite{wahls2015fast}.

\subsection{Inverse NFT}
The Inverse NFT (INFT) maps the given nonlinear spectrum to the corresponding pulse in time-domain.
For the special case of the spectrum without the continuous part, the 
Darboux Transformation can be applied to generate the
corresponding multi-soliton pulse~\cite{matveev1991darboux}. 
Algorithm~\ref{alg:DT2} shows the pseudo-code of the inverse transform, as described in \cite{aref2016control}.
 It generates an $N-$soliton $q\left(t\right)$ recursively
by adding a pair $\{\lambda_k,Q_d(\lambda_k)\}$ in each recursion. %\footnote{In Algorithm~\ref{alg:DT2}, $\lambda^*$ denotes the complex conjugate of $\lambda$. }.
The main advantage of this algorithm is that it is exact with a low computational complexity and it can be used to derive some properties of multi-soliton pulses. 

\begin{algorithm}

\SetKwInOut{Input}{Input}

\SetKwInOut{Output}{Output}

\Input{Discrete Spectrum $\{\lambda_k,Q_d(\lambda_k)\}$; $k=1,\dots,N$}

\Output{$N-$soliton waveform $q(t)$}

\BlankLine

\Begin{

\For{$k\leftarrow 1$ \KwTo $N$}{

$\rho_k^{(0)}(t)\longleftarrow\left(\frac{Q_d(\lambda_k)}{\lambda_k-\lambda_k^*}\prod_{m=1,m\ne k}^N \frac{\lambda_k-\lambda_m}{\lambda_k-\lambda_m^*}\right) e^{2j\lambda_k t}$\;

}

$q^{(0)} \longleftarrow 0$\;

\For{$k\leftarrow 1$ \KwTo $N$}{

$\rho(t)\longleftarrow \rho_k^{(k-1)}(t)$\;
\vspace*{-.5cm}
\begin{equation}\textstyle{
q^{(k)}(t)\longleftarrow q^{(k-1)}(t)+2j(\lambda_k-\lambda_k^*)\frac{\rho^*(t)}{1+|\rho(t)|^2}};
\label{eq:sig_update}
\end{equation}

\For{$m\leftarrow k+1$ \KwTo $N$}{
$\rho_m^{(k)}(t) \longleftarrow$
\vspace*{-.1cm}
\begin{equation}\textstyle{
\frac{(\lambda_m -\lambda_k)\rho_{m}^{(k-1)}(t)  +\frac{\lambda_k-\lambda_k^*}{1+|\rho(t)|^2} (\rho_{m}^{(k-1)}(t)-\rho(t))}
{\lambda_m -\lambda_k^*-\frac{\lambda_k-\lambda_k^*}{1+|\rho(t)|^2}\left(1 + \rho^*(t)\rho_{m}^{(k-1)}(t) \right)}};
\label{eq:rho_update}
\end{equation}

}

}

}

\footnotesize{($\lambda^*$ denotes the complex conjugate of $\lambda$)}

\caption{INFT from Darboux Transform\label{alg:DT2} \cite{aref2016control}}

\end{algorithm}

\vspace*{-.1cm}
\subsection{Definition of Pulse Duration and Bandwidth}\label{sec:def}
In this paper, we consider an $N-$soliton with the eigenvalues on the imaginary axis, i.e. $\{\lambda_k=j\sigma_k\}_{k=1}^N$ and $\sigma_k\in\mathbb{R}^+$. 
Without loss of generality, we assume that $\sigma_k<\sigma_{k+1}$.
As such an $N-$soliton propagates along the fiber, the pulse does not disperse and the pulse shape can be repeated periodically. 

An $N-$soliton pulse has unbounded support and
exponentially decreasing tails in time and (linear) frequency domain. As this pulse is transformed according to the NLSE, e.g. propagation along the ideal optical fiber, its shape can drastically 
change as all $Q_d(\lambda_k;z)$ are evolved in $z$.
Despite of nontrivial pulse variation
and various peak powers, the energy of the pulse
remains fixed and equal to $E_\mathrm{total}=4\sum_{k=1}^N\text{Im}\{\lambda_k\}$.

As a result, the pulse-duration and the bandwidth
of a multi-soliton pulse are well-defined if they are characterized in terms of energy:
the pulse duration $T_w$ (and bandwidth $B_w$, respectively) is defined as the smallest interval 
(frequency band) containing $E_\mathrm{trunc}=(1-\varepsilon)E_\mathrm{total}$ of the soliton energy.
Note that truncation causes small perturbations of eigenvalues. In practical applications, 
the perturbations become even larger
due to inter-symbol-interference (ISI) 
when a train of truncated soliton pulses is used for fiber optical communication.  
Thus, 
there is a trade-off:
$\varepsilon$ must be kept small such that the truncation causes only small perturbations,
but large enough to have a relatively small time-bandwidth product. 

Note that truncating a signal in time-domain may slightly change its linear Fourier spectrum in practice. For simplicity, we however computed $T_w$ and $B_w$ with respect to the original pulse as the difference is negligible for $\varepsilon\ll 1$.

\section{Symmetric Multi-Soliton Pulses}\label{sec:Sym_pulses}
In this section, we address the special family of multi-soliton pulses 
which are symmetric in time domain.
An application of such solitons for optical fiber transmission is studied in \cite{aref2015experimental} where the
symmetric 2-solitons are used for data modulation.
\begin{theorem}\label{thm:}
Let $\Omega=\{j\sigma_1,j\sigma_2,\dots,j\sigma_N\}$ be the set of eigenvalues on the imaginary axis
where $\sigma_k\in \mathbb{R}^+$, for $1\leq k\leq N$.
The corresponding $N-$soliton $q(t)$ is a symmetric pulse, i.e. $q(t)=q(-t)$,
and keeps this property during the propagation in $z$,
if and only if the spectral amplitudes are chosen as
\begin{equation}
	\left|Q_{d,\mathrm{sym}}\left(j\sigma_k\right)\right|=2\sigma_k \prod_{m=1;m\neq k}^N \left|\frac{\sigma_k+\sigma_m}{\sigma_k-\sigma_m}\right|. \label{Qd_abs_symmetric}
\end{equation}
\label{Theorem1:sym}
\end{theorem}
\textit{Sketch of Proof}. The proof is based on Algorithm~\ref{alg:DT2} with the following steps: (i) $g(t)=\frac{\rho^*(t)}{1+\vert \rho(t)\vert^2}$ is symmetric, if
\begin{equation}\label{eq:prop}
\rho^*(-t)\rho(t)=1.
\end{equation}
(ii) The update rule \eqref{eq:rho_update}
preserves the property \eqref{eq:prop}: if $\rho(t)$ and $\rho_m^{(k-1)}(t)$ satisfy \eqref{eq:prop}, then
$\rho_m^{(k)}(t)$ will satisfy \eqref{eq:prop} as well. \\
(iii) Because of \eqref{Qd_abs_symmetric}, $\rho_k^{(0)}(t)$ satisfies \eqref{eq:prop} for all $k$. \\
(iv) Using induction, $\rho_m^{(k)}$ satisfies \eqref{eq:prop} for all $m$ and $k$. 
\\(v) According to
\eqref{eq:sig_update} and step (i), $q(t)$ is symmetric. \qed

It is already mentioned in \cite{Haus1985} that \eqref{Qd_abs_symmetric} leads to a symmetric multi-soliton in amplitude. Theorem~\ref{Theorem1:sym} implies that \eqref{Qd_abs_symmetric} 
is not only sufficient but also necessary to have $q(t)=q(-t)$. 

As it is shown in the next section,
we \emph{numerically observe} that these symmetric pulses have the smallest pulse duration\footnote{It is correct when $\varepsilon$ is small enough.} among all solitons with
the same set of eigenvalues $\Omega$ (but different $\left|Q_d(\lambda_k)\right|$).
Assuming $\sigma_1=\min_k \left\{\sigma_k\right\}$, this minimum pulse-duration can be well approximated by
\begin{multline}
	T_\mathrm{sym}(\varepsilon)\approx \frac{1}{2\sigma_1}\left(2\sum_{m=2}^{N}\ln\left(\frac{\sigma_m+\sigma_1}{\sigma_m-\sigma_1}\right)\right.
	\\
	\left.  +\ln\left(\frac{2}{\varepsilon}\right)-\ln\left(\frac{\sum_{m=1}^{N}\sigma_m}{\sigma_1}\right)\right),
	 \label{T_symmetric}
\end{multline}
where $\varepsilon$ is defined earlier as the energy threshold. The approximation becomes tight as $\varepsilon\to 0$ and is only valid if $\varepsilon\ll \sigma_1/\sum_{m=1}^{N}{\sigma_m}$.
Verification of \eqref{T_symmetric} follows readily by describing an $N-$soliton by the sum of $N$ terms according to \eqref{eq:sig_update}, and showing that in the limit $|t|\to \infty$,
the dominant term behaves as ${\rm sech}(2\sigma_1 (|t|-t_0))$ for some $t_0$ and all other terms decay exponentially faster.

\section{Time-Bandwidth Product}\label{sec:tb}
Consider the transmission of an $N-$soliton with eigenvalues 
$\{j\sigma_k\}_{k=1}^N$ over an ideal fiber of length $z_L$. 
Each spectral amplitude $Q_d(j\sigma_k;z)=|Q_d(j\sigma_k;z)|\exp(j\phi_k(z))$ 
is transformed along the fiber according to \eqref{eq:qd_evol}. Equivalently,
\begin{align*}
|Q_d(j\sigma_k;z)|&=|Q_d(j\sigma_k;z=0)|\\
\phi_k(z)&=\phi_k(0)+4\sigma_k^2z
\end{align*}
for $z\leq z_L$. It means that $\phi_k(z)$ changes with a distinct speed proportional to 
$\sigma_k^2$. Different phase combinations correspond to different soliton pulse shapes with generally different pulse-duration and bandwidth. It implies that $T_w$ and $B_w$ of a pulse are changing along the transmission.
Furthermore, if the $\phi_k(0)$ are independently modulated for each eigenvalue with a constellation of size $M$, e.g. $M-$PSK, this results in $M^N$ initial phase combinations ($N\log_2(M)$ bits per soliton) associated with different initial pulse shapes. Such transmission scenarios are demonstrated experimentally for $M=4$, $N=2$~\cite{aref2015experimental} and 
$N=7$~\cite{Buelow20167eigenvalues}.  
To avoid a considerable ISI between neighboring pulses in a train of $N$-solitons for transmission in time or frequency, we should consider $T_w$ and $B_w$ larger than their respective maximum along the link.

For a given set of eigenvalues and fixed $|Q_d(j\sigma_k;z=0)|$, the maxima depend on $M^N$ initial phase combinations and the fiber length $z_L$.
To avoid these constraints, we maximize $T_w$ and $B_w$ over all possible phase combinations:
\begin{align*}
T_\mathrm{max}=\max_{\phi_k, 1\leq k\leq N} T_w \text{ and }
B_\mathrm{max}=\max_{\phi_k, 1\leq k\leq N} B_w.
\end{align*}
These quantities occur in the worst case 
but can be reached in their vicinity when $N$ is small, e.g. 2 or 3, or when $M$ is very large, or the transmission length $z_L$ is large enough. 

In the rest of this section, we address the following fundamental questions:
\begin{itemize}
\item[(i)] How do $T_{\rm max}$ and $B_{\rm max}$ change in terms of $\{j\sigma_k\}_{k=1}^N$ and $\{|Q_d(j\sigma_k)|\}_{k=1}^N$?
\item[(ii)] What is the smallest time-bandwidth product for a given $N$, i.e.
\begin{equation*}
(T_{\rm max}B_{\rm max})^\star=\min_{\sigma_k,1\leq k\leq N}\min_{|Q_d(j\sigma_k)|,1\leq k\leq N} 
T_{\rm max}B_{\rm max}
\end{equation*}
and which is the optimal choice for 
$\{j\sigma_k^\star\}_{k=1}^N$
and 
$\{|Q_d^\star(j\sigma_k^\star)|\}_{k=1}^N$.
\end{itemize}

The following properties preserving the time-bandwidth product decrease the number of parameters to optimize:
\\
(i) If $q(t)$ has eigenvalues $\{j\sigma_k\}_{k=1}^N$, 
then $1/\sigma_1 \cdot q(t/\sigma_1)$ will have eigenvalues $\{j\frac{\sigma_k}{\sigma_1}\}_{k=1}^N$ with the same time-bandwidth product.
It implies that $T_\mathrm{max}B_\mathrm{max}$ only 
depends on the $N-1$ eigenvalue ratios $\sigma_k/\sigma_1$.\\
(ii) If $\{\phi_k\}_{k=1}^N$ corresponds to $q(t)$, then $\{\phi_k-\phi_1\}_{k=1}^N$
corresponds to $q(t)\exp(j\phi_1)$. Thus, we assume $\phi_1=0$.\\
(iii) Instead of directly optimizing $\{|Q_d(j\sigma_k)|\}_{k=1}^N$, it is equivalent
to optimize $\eta_k>0$ defined by
\begin{equation*}
|Q_d(j\sigma_k)|= \eta_k |Q_{d,\mathrm{sym}}(j\sigma_k)|.
\end{equation*}
Using $\{\eta_k\}_{k=1}^N$ has two advantages. The first one is the generalization of Theorem~\ref{thm:}. If $\{\eta_k\}_{k=1}^N$ corresponds to $q(t)$, then $\{1/\eta_k\}_{k=1}^N$ corresponds to
$q(-t)$. The proof is similar to the one of Theorem~\ref{thm:}. Moreover, $\{e^{-2\sigma_kt_0}\eta_k\}_{k=1}^N$ corresponds to $q(t+t_0)$. Thus, it suffices to assume $\eta_1=1$ and 
$\eta_2\in(0,1]$.

\subsection{Optimization of Spectral Amplitudes}

Consider a given set of eigenvalues $\Omega=\{j\sigma_k\}_{k=1}^N$.
We want to optimize $\{\eta_k\}_{k=2}^N$
to minimize $T_{\rm max}B_{\rm max}$. Recall that $\{|Q_d(j\sigma_k)|\}_{k=1}^N$, and thus $\{\eta_k\}_{k=1}^N$ do not change along $z$.

We present the optimization method for $N=2$. 
In this case, there are two parameters to optimize: $\phi_2$ and $\eta_2\in(0,1]$.
Consider a given energy threshold $\varepsilon$.
For each chosen $\eta_2$, 
we find $T_{\rm max}(\varepsilon)$ and 
$B_{\rm max}(\varepsilon)$ by exhaustive search.
The phase $\phi_2\in[0,2\pi)$ is first quantized uniformly by 64 phases. At each phase, 
a 2-soliton is generated using Algorithm~\ref{alg:DT2}
and then $T_w(\varepsilon)$ and $B_w(\varepsilon)$
are computed.  To estimate $T_{\rm max}(\varepsilon)$, 
another round of search is performed with a finer resolution around the quantized phase with the largest $T_w(\varepsilon)$. Similarly, $B_{\rm max}(\varepsilon)$ is estimated.

\begin{figure}[tbh]
\centering
\begin{tikzpicture}[baseline=(current axis.south)]
\begin{semilogxaxis}[ticklabel style={font=\footnotesize},width=0.5\textwidth,height=0.2\textheight,xmin=10^-5,xmax=1,
xtick={0.00001,0.0001,0.001,0.01,0.1,1},xticklabels={-5,-4,-3,-2,-1,0},ymin=5,ymax=33,ylabel style={yshift=-1em},label style={font=\small},ylabel={$T_\mathrm{max}$},xlabel={$\log\left(\eta_2\right)$},x label style={yshift=0.8em},title style={font=\small},legend style={at={(1.01,-0.225\textheight)},
    anchor=east,font=\scriptsize,legend columns=4,overlay}
    ]
    \addplot[forget plot,thick,myblue,width=\linewidth,each nth point={2}]table {my_figures/Data/T-1.dat};
    \addplot[forget plot,thick,myred,width=\linewidth]table {my_figures/Data/T-2.dat};
    \addplot[forget plot,thick,mypurple,width=\linewidth]table {my_figures/Data/T-3.dat};
    \addplot[forget plot,thick,mygreen,width=\linewidth]table {my_figures/Data/T-4.dat};

\addlegendimage{line legend,mypurple}
\addlegendentry{$\varepsilon=10^{-3}$}
\addlegendimage{line legend,mygreen}
\addlegendentry{$\varepsilon=10^{-4}$}
\addlegendimage{line legend,myred}
\addlegendentry{$\varepsilon=10^{-6}$}
\addlegendimage{line legend,myblue}
\addlegendentry{$\varepsilon=10^{-10}$}    
    
\end{semilogxaxis}
\node [below=0.5cm, align=flush center,text width=1cm] at (0.5,1) {(a)};

\end{tikzpicture}
\vspace{-1mm}
\begin{tikzpicture}[baseline=(current axis.south)]
\begin{semilogxaxis}[ticklabel style={font=\footnotesize},width=0.5\textwidth,height=0.2\textheight,xmin=10^-5,xmax=1, xtick={0.00001,0.0001,0.001,0.01,0.1,1},xticklabels={-5,-4,-3,-2,-1,0}, ymin=0,ymax=8,y label style={yshift=-1em},label style={font=\small},ylabel={$B_\mathrm{max},B_\mathrm{min}$}
,xlabel={$\log\left(\eta_2\right)$},x label style={yshift=0.8em},title style={font=\small},legend style={legend pos=north west,font=\footnotesize}]
    \addplot[forget plot,thick,dashed,myblue,width=\linewidth]table {my_figures/Data/B-1.dat};
    \addplot[forget plot,thick,myblue,width=\linewidth]table {my_figures/Data/B-2.dat};
    \addplot[forget plot,thick,dashed,myred,width=\linewidth]table {my_figures/Data/B-3.dat};
    \addplot[forget plot,thick,myred,width=\linewidth]table {my_figures/Data/B-4.dat};
    \addplot[forget plot,thick,dashed,mypurple,width=\linewidth]table {my_figures/Data/B-5.dat};
    \addplot[forget plot,thick,mypurple,width=\linewidth]table {my_figures/Data/B-6.dat};
    \addplot[forget plot,thick,dashed,mygreen,width=\linewidth]table {my_figures/Data/B-7.dat};
    \addplot[forget plot,thick,mygreen,width=\linewidth]table {my_figures/Data/B-8.dat};

\addlegendimage{line legend,gray}
\addlegendentry{$B_\mathrm{max}$}
\addlegendimage{line legend,dashed,gray}
\addlegendentry{$B_\mathrm{min}$}

\end{semilogxaxis}
\node [below=0.5cm, align=flush center,text width=1cm] at (0.5,1) {(b)};

\end{tikzpicture}
\vspace{6mm}
\caption{(a) Pulse duration $T_\mathrm{max}$ and (b) bandwidth $B_\mathrm{max/min}$ for 2-soliton pulse ($\lambda_1=0.5j$, $\lambda_2=1j$) when maximized (minimized) over all phase combinations of spectral amplitudes}
\label{fig:T_B}
\vspace{-5mm}
\end{figure}
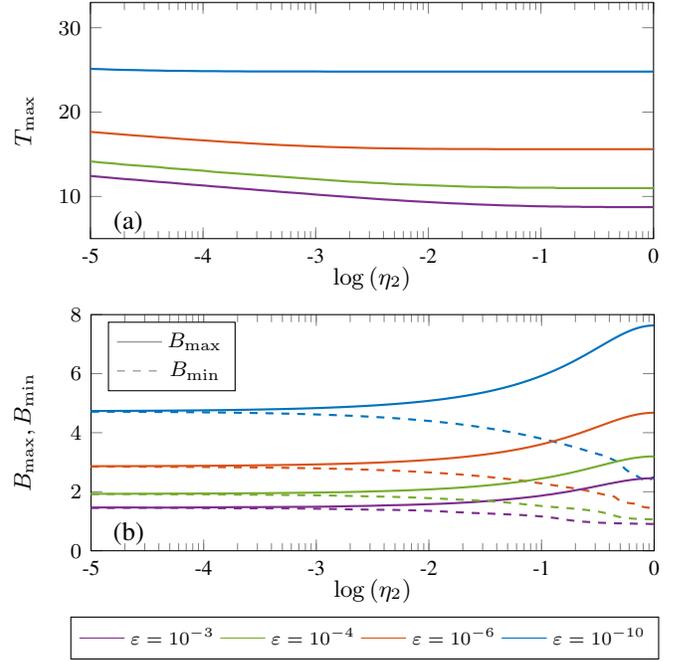

Fig.~\ref{fig:T_B} illustrates $T_{\rm max}(\varepsilon)$ and $B_{\rm max}(\varepsilon)$ 
in terms of $\log(\eta_2)$ for different energy 
thresholds $\varepsilon$ when  
$\Omega=\{\frac{1}{2}j,1j\}$.
We also depict $B_\mathrm{min}(\varepsilon)$, the minimum bandwidth of 2-soliton pulses with a given $\eta_2$ and various $\phi_2$. Fig.~\ref{fig:T_B}
indicates the following features that we observed for any pairs of $\{j\sigma_1,j\sigma_2\}$.
 
We can see that for any $\varepsilon$, the smallest $T_\mathrm{max}$ is attained at $\eta_2=1$ ($\log(\eta_2)=0$) which corresponds to the symmetric 2-soliton defined in 
Sec.~\ref{sec:Sym_pulses}. We also observe that $B_\mathrm{max}$ reaches the largest value at $\eta_2=1$ while $B_\mathrm{min}$
reaches its minimum.

As $\log(\eta_2)$ decreases, $T_\mathrm{max}$ increases gradually up to some point and then it linearly increases in $|\log(\eta_2)|$. The behaviour
of $B_\mathrm{max}$ is the opposite. It decreases very fast in $|\log(\eta_2)|$ up to some $\eta_2$ and then converges slowly to the bandwidth defined by the 1-soliton spectrum with $\lambda=j\sigma_2$. In fact, we have two separate 1-solitons without any interaction when $\eta_2=0$.
As $\eta_2$ increases to $1$, the distance between these two 1-solitons decreases, resulting in more nonlinear interaction but smaller $T_\mathrm{max}$. 
The largest $B_\mathrm{max}-B_\mathrm{min}$ at $\eta_2=1$ indicates
the largest amount of interaction.

The above features seem general for $N-$solitons.
In particular, $T_\mathrm{max}$ becomes minimum if 
the $N-$soliton is symmetric. Moreover,
$B_\mathrm{max}$ can be lower-bounded by 
\begin{equation*}
B_\mathrm{sep}\left(\varepsilon\right)=\frac{2\sigma_N}{\pi^2}\left(\ln\left(\frac{2}{\varepsilon}\right)-\ln\left(\frac{\sum_{k=1}^{N}\sigma_k}{\sigma_N}\right)\right)
\end{equation*}
with $\sigma_N=\max_k \left\{\sigma_k\right\}$. The bound becomes tight when an $N-$soliton is the linear superposition of $N$ separate 1-solitons.

We performed such a numerical optimization for $N=2$ and $N=3$ and for different $\{\frac{\sigma_k}{\sigma_1}\}_{k=2}^N$. For each $\varepsilon$, we found the optimal $\{\eta_k^\star\}_{k=2}^N$ with the smallest $T_\mathrm{max}(\varepsilon)B_\mathrm{max}(\varepsilon)$.

\subsection{Optimization of Eigenvalues}

In general, an $N-$soliton has a larger $T_\mathrm{max}B_\mathrm{max}$ than a $1-$soliton but it has also $N$ times, e.g. $Q_d(j\sigma_k)$, more dimensions for encoding data.   
To have a fair comparison, 
we use a notion of ``time-bandwidth product per eigenvalue'' defined as
\begin{equation*}
\overline{T \! \cdot \! B}_{N}(\{\frac{\sigma_k}{\sigma_1}\}_{k=2}^N)=\frac{1}{N}T_\mathrm{max}B_\mathrm{max}(\{\frac{\sigma_k}{\sigma_1}\}_{k=2}^N,\{\eta_k^\star\}_{k=2}^N)
\end{equation*}
where $T_\mathrm{max}B_\mathrm{max}$ is already optimized in terms of $\{\eta_k\}$, seperately for each eigenvalue combination.
This is an important parameter as the spectral efficiency will be 
$\mathcal{O}(1/\overline{T \! \cdot \! B}_{N})$.
For a 1-soliton with ``sech'' shape in time and frequency domain, we have 
\begin{equation*}
\overline{T \! \cdot \! B}_{1}=T_w(\varepsilon)B_w(\varepsilon)=\pi^{-2}\ln^2(2/\varepsilon),
\end{equation*}
where $\varepsilon$ is the energy threshold defined in Section~\ref{sec:def}.

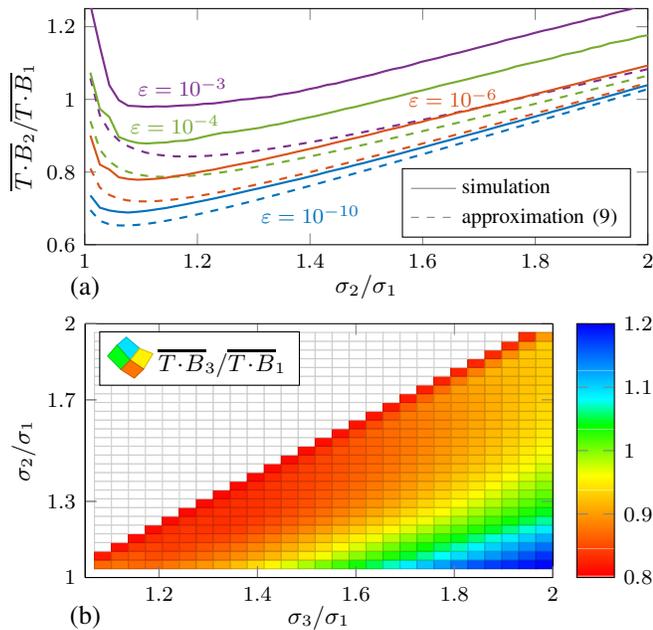
\begin{figure}[tb]
\centering
\begin{tikzpicture}[baseline=(current axis.south)]
\begin{axis}[ticklabel style={font= \footnotesize},width=0.5\textwidth,height=0.2\textheight,xmin=1,xmax=2,ymin=0.6,ymax=1.25,y label style={yshift=-1.2em},label style={font=\small},ylabel=$\overline{T \! \cdot \! B}_{2}/\overline{T \! \cdot \! B}_{1}$,xlabel={$\sigma_2/\sigma_1$},x label style={yshift=0.7em},legend style={font=\footnotesize},legend pos=south east,legend cell align=left]
    \addplot[forget plot,thick,dashed,myblue,width=\linewidth]table[y expr=1/\thisrowno{1}] {my_figures/Data/SE_gain-6.dat};
    \addplot[forget plot,thick,myblue,width=\linewidth]table[y expr=1/\thisrowno{1}] {my_figures/Data/SE_gain-9.dat};
    \addplot[forget plot,thick,dashed,mypurple,width=\linewidth]table[y expr=1/\thisrowno{1}] {my_figures/Data/SE_gain-1.dat};
     \addplot[forget plot,thick,mypurple,width=\linewidth]table[y expr=1/\thisrowno{1}] {my_figures/Data/SE_gain-12.dat};      
    \addplot[forget plot,thick,dashed,mygreen,width=\linewidth]table[y expr=1/\thisrowno{1}] {my_figures/Data/SE_gain-2.dat};
    \addplot[forget plot,thick,mygreen,width=\linewidth]table[y expr=1/\thisrowno{1}] {my_figures/Data/SE_gain-11.dat};
    \addplot[forget plot,thick,dashed,myred,width=\linewidth]table[y expr=1/\thisrowno{1}] {my_figures/Data/SE_gain-4.dat};
    \addplot[forget plot,thick,myred,width=\linewidth]table[y expr=1/\thisrowno{1}] {my_figures/Data/SE_gain-8.dat};
  
\addlegendimage{line legend,gray}
\addlegendentry{simulation}
\addlegendimage{line legend,dashed,gray}
\addlegendentry{approximation~\eqref{TB_approx}}

\end{axis}

\node [below=0.4cm, align=flush center,text width=1cm] at (0,0.1) {(a)};

\node[color=mypurple,font=\footnotesize] at (1.3,2.1) {$\varepsilon=10^{-3}$};
\node[color=mygreen,font=\footnotesize] at (1.2,1.6) {$\varepsilon=10^{-4}$};
\node[color=myred,font=\footnotesize] at (4.9,1.95) {$\varepsilon=10^{-6}$};
\node[color=myblue,font=\footnotesize] at (3,0.4) {$\varepsilon=10^{-10}$};

\end{tikzpicture}
\vspace{-2mm}
\begin{tikzpicture}[baseline=(current axis.south)]
\begin{axis}[ticklabel style={font=\footnotesize},width=0.43\textwidth,height=0.21\textheight,xmin=1.05,xmax=2,ymin=1,ymax=2,y label style={yshift=-0.25em},label style={font=\small},zlabel=$\frac{SE_2}{SE_1}$,xlabel={$\sigma_3/\sigma_1$},x label style={yshift=0.5em},ylabel={$\sigma_2/\sigma_1$},ytick={1,1.3,1.7,2},title style={font=\small},legend pos=north west,legend style={font=\small},view={0}{90},colormap={custom}{color(0)=(white) color(0.01)=(red) color(0.4)=(yellow) color(0.6)=(green) color(0.7)=(cyan) color(1)=(blue)},colorbar,point meta min=0.8,point meta max=1.2,colorbar style={xshift=-0.5em}]
    \addplot3[surf,mesh/ordering=y varies,mesh/rows=28]table {my_figures/Data/SE_gain_inverse_3.dat};   

\addlegendentry{$\overline{T \! \cdot \! B}_{3}/\overline{T \! \cdot \! B}_{1}$}    
     
\end{axis}

\node [below=0.35cm, align=flush center,text width=1cm] at (0,0.1) {(b)};
\end{tikzpicture}
\caption{Gain of time-bandwidth product per eigenvalue of (a) second and (b) third order solitons with eigenvalues $j\sigma_k$ in relation to first order pulses}
\label{fig:SE}
\vspace{-5mm}
\end{figure}

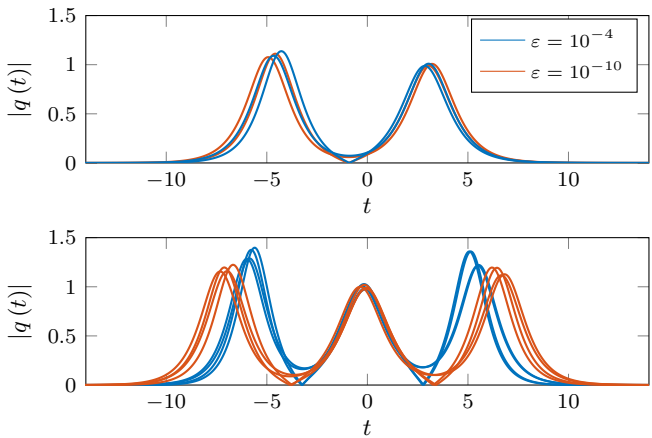
\begin{figure}[tbh]
\centering
\begin{flushleft}
\begin{tikzpicture}[baseline=(current axis.south)]
\begin{axis}[ticklabel style={font=\footnotesize},width=0.5\textwidth,height=0.15\textheight,xmin=-14,xmax=14,ymin=0,ymax=1.5,y label style={yshift=-1em},label style={font=\small},ylabel={$\left|q\left(t\right)\right|$},xlabel={$t$},x label style={yshift=0.5em},legend style={font=\scriptsize},legend cell align=left,title style={font=\small}
]     
    
    \addplot[forget plot,thick,myred,width=\linewidth]table {my_figures/Data/signal_opt_t_epsilon-1.dat};
    \addplot[forget plot,thick,myred,width=\linewidth]table {my_figures/Data/signal_opt_t_epsilon-2.dat};
    \addplot[forget plot,thick,myblue,width=\linewidth]table {my_figures/Data/signal_opt_t_epsilon-5.dat};
    \addplot[forget plot,thick,myblue,width=\linewidth]table {my_figures/Data/signal_opt_t_epsilon-6.dat};

\addlegendimage{line legend,myblue}
\addlegendentry{$\varepsilon=10^{-4}$}
\addlegendimage{line legend,myred}
\addlegendentry{$\varepsilon=10^{-10}$}    
    
\end{axis}
\end{tikzpicture}
\begin{tikzpicture}[baseline=(current axis.south)]
\begin{axis}[ticklabel style={font=\footnotesize},width=0.5\textwidth,height=0.15\textheight,xmin=-14,xmax=14,ymin=0,ymax=1.5,y label style={yshift=-1em},label style={font=\small},ylabel={$\left|q\left(t\right)\right|$},xlabel={$t$},x label style={yshift=0.5em},legend style={font=\tiny},title style={font=\small}]

    \addplot[forget plot,thick,myblue,width=\linewidth]table {my_figures/Data/signal3_opt_t_epsilon-1.dat};
    \addplot[forget plot,thick,myblue,width=\linewidth]table {my_figures/Data/signal3_opt_t_epsilon-2.dat};
    \addplot[forget plot,thick,myblue,width=\linewidth]table {my_figures/Data/signal3_opt_t_epsilon-3.dat};
    \addplot[forget plot,thick,myblue,width=\linewidth]table {my_figures/Data/signal3_opt_t_epsilon-4.dat};   
    \addplot[forget plot,thick,myred,width=\linewidth]table {my_figures/Data/signal3_opt_t_epsilon-5.dat};
    \addplot[forget plot,thick,myred,width=\linewidth]table {my_figures/Data/signal3_opt_t_epsilon-6.dat};  
    \addplot[forget plot,thick,myred,width=\linewidth]table {my_figures/Data/signal3_opt_t_epsilon-7.dat};
    \addplot[forget plot,thick,myred,width=\linewidth]table {my_figures/Data/signal3_opt_t_epsilon-8.dat};
             
\end{axis}
\end{tikzpicture}
\end{flushleft}
\vspace{-2mm}
\caption{Time domain signal of optimum second and third order soliton pulse for different phase combinations of the spectral amplitudes (same color)}
\label{fig:signal_time_domain}
\end{figure}

For $N=2$ and $N=3$, we numerically optimized $T_\mathrm{max}(\varepsilon)B_\mathrm{max}(\varepsilon)$
for different values of $\{\frac{\sigma_k}{\sigma_1}\}_{k=2}^N$ and $\{\eta_k\}_{k=2}^N$. Fig.~\ref{fig:SE}-$(a)$ shows the numerical optimization of
$\overline{T \! \cdot \! B}_{2}$ in terms of $\sigma_2/\sigma_1$ for different choices of $\varepsilon$ where the best $\{\eta_k^\star\}_{k=2}^N$ were chosen for each eigenvalue ratio.
We normalized $\overline{T \! \cdot \! B}_{2}$ by $\overline{T \! \cdot \! B}_{1}$ to see how much the ``time-bandwidth product per eigenvalue'' can be decreased. 
Fig.~\ref{fig:SE}-$(b)$ shows a similar numerical optimization for $N=3$ and $\varepsilon=10^{-4}$.
We have the following observations:\\
(i) $\overline{T \! \cdot \! B}_{N}$ is sensitive to the choice of eigenvalues.
For instance, equidistant eigenvalues, i.e. $\sigma_k=k\sigma_1$, are a bad choice in terms of
spectral efficiency.\\ 
(ii) The ratio $\overline{T \! \cdot \! B}_{N}/\overline{T \! \cdot \! B}_{1}$ gets smaller as $\varepsilon$ vanishes. The intuitive reason is that as $\varepsilon\to 0$, we get $T_{\rm max}\approx\frac{1}{2\sigma_1}\ln(\frac{2}{\varepsilon})$ (see \eqref{T_symmetric}) which is the pulse-duration of the 1-soliton.
\\
(iii) For a practical value of $\varepsilon\sim 10^{-4}-10^{-3}$, 
$\overline{T \! \cdot \! B}_{N}$ decreases very slowly in $N$. Moreover, the optimal $\sigma_k^{\star}$ are close. This can make the detection challenging in presence of noise. For $\varepsilon=10^{-4}$,
 \begin{align*}
 \overline{T \! \cdot \! B}_{2}/\overline{T \! \cdot \! B}_{1}=0.87  &\text{ for }\sigma_2^\star/\sigma_1^\star=1.11\\
 \overline{T \! \cdot \! B}_{3}/\overline{T \! \cdot \! B}_{1}=0.83 &\text{ for }\sigma_2^\star/\sigma_1^\star=1.28,
 \sigma_3^\star/\sigma_1^\star=1.35 
 \end{align*}
(iv) Choosing the above optimal $\{\sigma_k^\star/\sigma_1^\star\}$, and the optimal
 $\{|Q_d^\star(\sigma_k^\star)\}$, the resulting solitons for $N=2,3$ are shown in Fig.~\ref{fig:signal_time_domain} for different phase combinations and two energy thresholds $\varepsilon$. This figure gives some guidelines for a larger $N$: the optimal $N-$soliton has eigenvalues close to each other and significantly seperated pulse centers, why the optimum pulse looks similar to a train of 1-solitons with eigenvalues close to each other.
The pulse centers should be close to minimize $T_{\rm max}$ but not too close to avoid a large interaction which comes along with a growth of $B_{\rm max}$.

\vspace{-2mm}

For $\varepsilon\ll 1$, an estimate on $\overline{T \! \cdot \! B}_{N}$ at optimal $\{\eta_k^*\}$ can be given by \eqref{TB_approx}, where $T_{\rm max}$ and $B_{\rm max}$ are estimated by $T_{\rm sym}(\varepsilon)$ and $B_{\rm sep}(\varepsilon)$, respectively (see Fig.~\ref{fig:T_B}).
 
\begin{equation}\label{TB_approx}
\overline{T \! \cdot \! B}_{N}\approx
\frac{T_{\rm sym}(\varepsilon)
B_\mathrm{sep}\left(\varepsilon\right)}{N},
\end{equation}
 For the second order case, these approximations for various $\varepsilon$ are plotted in Fig. \ref{fig:SE}-$(a)$ by dashed lines. We see that the approximation becomes
better for small $\varepsilon$. This approximation can be used to predict $\overline{T \! \cdot \! B}_{N}$ for a large $N$.

\section{Conclusion}\label{sec:con}

We studied the evolution of the pulse-duration 
and the bandwidth of $N-$soliton pulses along the optical fiber. We focused on solitons with eigenvalues located on the imaginary axis. The class of symmetric soliton pulses was introduced and an analytical approximation of their pulse-duration was derived. 

The phase of the spectral amplitudes was assumed to be used for modulation while their magnitudes were kept fixed. We numerically optimized the location of eigenvalues and the magnitudes of spectral amplitudes for $2-$ and $3-$solitons in order to minimize the time-bandwidth product. It can be observed that 
the time-bandwidth product per eigenvalue improves in the soliton order $N$, but very slowly. Another observation is, that the optimal $N-$soliton pulse looks similar to a train of first-order pulses.

There are some remarks about our optimization.
As an $N-$soliton propagates, the phases of the spectral amplitudes change with different speeds. We assumed that all possible combinations of phases occur during transmission. This is the worst case scenario which is likely to happen for $N=2$ and $N=3$ but becomes less probable for large $N$. Moreover, the same magnitudes of spectral amplitudes are used for any phase combination while they can be tuned according to the phases. Without these assumptions, the time-bandwidth product 
will decrease. However, it becomes harder to estimate as there are many more parameters to optimize.

\bibliographystyle{IEEEtran}
\bibliography{references_nft}

\end{document}